\documentclass[aps,prc,preprint,groupedaddress,showpacs,eqsecnum]{revtex4}
\usepackage{graphicx}
\usepackage{dcolumn}
\usepackage{bm}

\newcommand{\taN}{$^{180}$Ta}

\newcommand{\taisob}{$^{180}$Ta$^{m}$}

\newcommand{\gpro}{$\gamma$-process}
\newcommand{\nupro}{$\nu$-process}

\newcommand{\hiK}{high-$K$}
\newcommand{\loK}{low-$K$}

\begin{document}

\preprint{Published in Phys. Rev. C 82, 058801(R)(2010)}

\title{
Re-analysis of the ($J =5$) state at 592\,keV in $^{180}$Ta and its role
in the $\nu$-process nucleosynthesis of $^{180}$Ta in supernovae
}

\author{T.\ Hayakawa}
\email[E-mail: ]{hayakawa.takehito@jaea.go.jp}
\affiliation{Quantum Beam Science Directorate, Japan Atomic Energy Agency,
  Shirakara-Shirane 2-4, Tokai-mura, Ibaraki 319-1195, Japan}
\affiliation{National Astronomical Observatory, Osawa, Mitaka, Tokyo 181-8588,
  Japan}

\author{P.\ Mohr}
%
%
\affiliation{
Diakoniekrankenhaus Schw\"abisch Hall, D-74523 Schw\"abisch Hall,
Germany}
\affiliation{
Institute of Nuclear Research (ATOMKI), H-4001 Debrecen, Hungary}

\author{T.\ Kajino}
\affiliation{National Astronomical Observatory, Osawa, Mitaka, Tokyo 181-8588,
  Japan}
\affiliation{Department of Astronomy, Graduate School of Science, University
  of Tokyo, Tokyo 113-0033, Japan}

\author{S.\ Chiba}
\affiliation{Advanced Science Research, Japan Atomic Energy Agency,
  Shirakara-Shirane 2-4, Tokai-mura, Ibaraki 319-1195, Japan}

\author{G.\ J.\ Mathews}
\affiliation{Center for Astrophysics, Department of Physics, University of
  Notre Dame, Notre Dame, IN 46556}

\date{\today}

\begin{abstract}
We analyze the production and freeze-out of the isomer \taisob\ in
the $\nu$-process. We consider
the influence of a possible low-lying intermediate ($J = 5$)
state at 592\,keV using a transition width estimated from the measured half-life.
This more realistic width 
is much smaller than the previous estimate. 
We find that the 592\,keV state leads only to a small reduction of the residual
isomer population ratio from the previous result; i.e., considering this better  estimate for the transition width, the isomer population ratio changes from ${\cal{R}} = 0.39$
to ${\cal{R}} = 0.38$, whereas previously it was estimated that this transition could reduce the ratio to  
${\cal{R}} = 0.18$. 
This finding strengthens 
the evidence that $^{138}$La and $^{180}$Ta are coproduced by neutrino
nucleosynthesis with an electron neutrino temperature of $kT \approx 4$\,MeV.
\end{abstract}


\maketitle

The possible astrophysical origin of the rarest isotope in Nature, $^{180}$Ta$^m$,  has been speculated upon for the last 30 years \cite{Beer81,Yokoi83,Woo78,Woosley90}.  The problem with this isotope lies in the fact that it is bypassed in the normal nucleosynthesis processes for heavy nuclei.
The most promising production scenario in recent times 
involves neutrino-induced reactions on $^{180}$Hf and $^{181}$Ta  
during  core-collapse supernovae (i.e.~ the $\nu$ process) \cite{Woosley90,Heger05,Byelikov07}.
However,  previous calculations  could not satisfactorily explain  the solar abundance of $^{180}$Ta$^m$
relative to $^{16}$O (or $^{24}$Mg) because they lacked a realistic
calculation of  the residual population ratio of the isomer to the true ground state.  Knowing this ratio is crucial because the true ground state decays with a lifetime of only $8.15~h$, 
while the naturally occurring state is $^{180}$Ta$^m$ with a lifetime $> 10^{15}~y$ (see Fig.~1).  
The purpose of this paper is to demonstrate that by carefully   considering the limits on the lowest possible state which can connect the ground state and the isomer, it is possible to place a much more stringent constraint on the final residual ratio.

In the $\nu$ process scenario of interest here \taN\ is produced
at relatively high temperatures so that  a  thermal equilibrium exists between the long-lived
\hiK\ $9^-$ isomer at 77\,keV and the short-lived \loK\ $1^+$ ground
state.
During freeze-out the \hiK\ states and the \loK\ states become
decoupled because the thermal photon bath is no longer able to populate
higher-lying so-called intermediate states (IMSs) which decay to both the
\loK\ and the \hiK\ states of \taN. 

In order to clarify the residual ratio between the ground state and the isomer, in a recent paper \cite{Haya10} we studied the nucleosynthesis of \taisob\ in the
\nupro. The time-dependent evolution of the
residual isomeric ratio was carefully followed, see Eqs.~(1)--(4) in
\cite{Haya10}. Details of this evolution are independent of the production
mechanism and are thus also valid for the so-called \gpro\ that probably
occurs in type II supernova explosions \cite{Woo78,Arn03,Hayakawa04,Hayakawa08}.

In the previous study \cite{Haya10} 
we utilized experimental values for the transition strengths
and integrated cross sections of the known IMSs in \taN\ which have been measured by
photoactivation \cite{Bel99,Bel02}. The resulting isomeric residual population
ratio ${\cal{R}} = P_m/(P_g + P_m)$ was found to be ${\cal{R}} = 0.39 \pm
0.01$,
where the uncertainty was evaluated from the experimental errors 
of the energy width \cite{Bel02}.
This result was an improvement upon  an earlier determination  \cite{Mohr07} of ${\cal{R}} =0.35 \pm 0.04$ that was
 obtained from an estimate of the freeze-out temperature without following
the time-dependent evolution in detail.
We also found that the final result ${\cal{R}}$ = 0.39 $\pm$ 0.01 
was nearly independent of a number of astrophysical parameters such as
the progenitor mass, 
the supernova explosion energy, the neutrino energy spectrum, or the peak temperature of
the nucleosynthesis environment.  This independence arises because 
 \taN\ is completely thermalized in the initial environment where the  
 temperature is  high $T_9 > 0.62$ (where $T_9$ is the temperature in $10^9$\,K).
Also, although the final isomer residual ratio is determined from the  time-dependent
evolution during  freeze out, it is rather insensitive to the time scale for freezeout \cite{Haya10}.

It has not been possible until recently  to find the IMSs in \taN\ by $\gamma$-ray spectroscopy \cite{Dra98,Sai99,Dra00,Wen01}. 
However, the existence of IMSs has now been
clearly confirmed by photoactivation experiments (see \cite{Bel99,Bel02}, and
earlier experiments referenced therein). 
The lowest experimentally confirmed IMS is located 1.01\,MeV above the $9^-$
isomer at 77\,keV, leading to a total excitation energy of 1087\,keV with an
uncertainty of about 10\,keV.
Walker {\it et al.}\ \cite{Wal01}  
pointed out 
that the excitation energies of the lowest three IMSs  (1087,
1297,  and 1507\,keV) with experimental uncertainties of $10-30$\,keV \cite{Bel99,Bel02}
are "very close" to the energies of three states
which were measured by  $\gamma$-ray spectroscopy experiments  to be at 1076,
1277, and  1499\,keV.
The three excited states observed by the $\gamma$-ray spectroscopy
are members of a rotational band, whose band head is located at 592\,keV. 

This 592-keV state predominantly decays to a 4$^+$ state at 520 keV
with a measured half-life of $T_{1/2} = 16.1 \pm 1.9$\,ns.
It was suggested that the 592-keV state may decay to
the $7^+$ state at 357\,keV on the \hiK\ side via a $K$-allowed 235-keV $\gamma$-ray (see Fig.\ 1) \cite{Mohr07}. 
This state cannot be detected in photoactivation experiments
because there is no direct branch from the 592\,keV state to the $9^-$ isomer.
We estimated the
transition width of this state in our previous paper \cite{Haya10} using an
exponential extrapolation (see Fig.~4 in \cite{Haya10}).  
In this way we inferred  a transition width of $\Gamma_{\rm{tr}} =
(g_{\rm{IMS}}/g_{\rm{iso}}) \Gamma_{\rm{iso}} = 3$ $\times$ 10$^{-6}$ eV.  
This  leads to a significantly different isomer ratio of ${\cal{R}} = 0.18$ 
instead of ${\cal{R}} = 0.39$.
However, since there is no experimental evidence suggesting the existence a transition to the  IMS at 592 keV,
this smaller ${\cal{R}}$ was not adopted in \cite{Haya10}.
A transition width of $\Gamma_{\rm{tr}}$ =
3 ${\times} 10^{-6}$ eV  would imply a lifetime for this state of $T_{1/2}  < 0.15$ ns.   However,
as seen from Fig.~1, this state has a much longer measured  half-life  of $T_{1/2} = 16 \pm 1.6$ ns even if 100\% of the decay of this state is via the $E2$ transition, 
implying that our previous 
estimate of the width for this state was an overestimate by more than two orders of magnitude.

A small transition width of $\Gamma_{\rm{tr}} = 1.6 \times 10^{-10}$\,eV
for the 235-keV $\gamma$-ray was estimated with an assumption of the 1\,\% branch
for this $\gamma$-ray and the measured half-life of the 592-keV state \cite{Mohr07}.
This corresponds to
7.5 $\times$ 10$^{-3}$ Weisskopf units (W.u.)
Based upon these arguments, we have repeated our calculation of the time-dependent 
freeze-out of
\taN\ using the more realistic much smaller estimate for the transition width of
$\Gamma_{\rm{tr}} = 1.6 \times 10^{-10}$\,eV for the 592\,keV state \cite{Mohr07}. 
The resulting isomeric ratio is ${\cal{R}} = 0.38$, 
quite close to the value ${\cal{R}} =
0.39$ calculated from the experimentally known IMSs, 
but much larger than ${\cal{R}} = 0.18$ based upon 
the previous unrealistically large extrapolated  width of $\Gamma_{\rm{tr}} =
3$ $\times$ 10$^{-6}$ eV. 
This result is not surprising.  The transition rates
$\lambda^\ast$ between \hiK\ and \loK\ states in \taN\ of the 1087\,keV state
(experimentally confirmed) and of the 592\,keV (using the above estimated transition width) state are almost identical in
the freeze-out temperature region (see Fig.~\ref{fig:rates}).

The parity assignment of the 592-keV state is critical for the existence 
of the 235-keV transition.
Although a 5$^+$ assignment for 592 keV state is favored,
the parity has not been experimentally established \cite{Wu03}.
In the case of a $5^-$ assignment, again an E2 transition
to the \hiK\ side with $\Delta K = 2$ to the $7^-$ state at 463\,keV is possible.
However, the estimated transition
strength may be somewhat lower than in the case of the $5^+$ assignment
because of the smaller transition energy; 
this finally leads to isomeric ratios between ${\cal{R}} = 0.38$ and 0.39.
The influence of the
592\,keV state on the isomer residual ratio ${\cal{R}}$ remains negligibly small 
under any conditions.

Many $\gamma$-ray spectroscopy experiments \cite{Dra98,Sai99,Dra00,Wen01}
have been carried out but the
235-keV $\gamma$-ray or other decay branches of the 592\,keV state have
not yet been detected. 
These transitions may be strongly hindered. Such hindrances
are certainly
expected when there are configuration changes such as a change of
both neutron and proton orbitals \cite{Gallagher60}.
In fact large hindrance factors of 10$^{-2}$$\--$10$^{-6}$ W.u.\ 
have been measured for the $K$-allowed transitions in $^{180}$Ta \cite{Sai99}.
If it were experimentally confirmed that the upper limit of the decay branches 
is less than 1\%,
this would be evidence supporting the $\nu$ process origin of $^{180}$Ta.
In the $s$ process,
the 592\,keV state will be the
most important IMS at temperatures around $kT \approx 23$\,keV.
If the decay branches are experimentally confirmed,
the effective half-life of $^{180}$Ta decreases drastically and 
the survival of $^{180}$Ta in the $s$ process becomes difficult \cite{Mohr07}.
Therefore a measurement of the weak decay
branches of the 592\,keV state would be highly desirable although 
this would be a challenging experiment as in the case of $^{176}$Lu \cite{176Lu}.

We should also briefly mention that the influence of the surrounding plasma on
the nuclear transitions in \taN\ has been studied recently \cite{Gos10}. It
was found that within the present knowledge of IMSs  the surrounding plasma does
not have significant influence on the nucleosynthesis of \taN .

In conclusion, we have taken into account the claimed lowest IMS in \taN\ at
592\,keV for the time-dependent evolution of the isomeric ratio ${\cal{R}}$ in
the freeze-out of core-collapse supernova explosions. 
Using a $5^+$ assignment and a more realistic estimate 
\cite{Mohr07} for the transition width, we find that the isomeric ratio
decreases slightly from ${\cal{R}} = 0.39$ to ${\cal{R}} = 0.38$ for the
nucleosynthesis of \taN\ in $\gamma$- and $\nu$-processes in supernova
explosions. 
Another assignment $J^{\pi}$ = $5^-$ leads to even smaller
modifications of ${\cal{R}}$. 
Hence, even if the 592-keV state is indeed the
lowest IMS in \taN, 
its influence remains negligibly small 
for the freeze-out in the $\nu$- and $\gamma$-processes.
The main conclusion of the previous study \cite{Haya10} is thus
strengthened: the solar abundances of $^{138}$La and \taN\ relative to $^{16}$O
can be  systematically 
reproduced by neutrino nucleosynthesis and an electron neutrino temperature of
$kT \approx 4$\,MeV.
\\

This work has been supported in part by Grants-in-Aid for Scientific
Research (21340068, 20244035, 20105004) of Japan.   
Work at the University of Notre Dame (G.J.M.) supported
by the U.S. Department of Energy under 
Nuclear Theory Grant DE-FG02-95-ER40934.
This work was also supported by OTKA (NN83261).

\begin{figure*}[thbp] 
  \centering
  \includegraphics[scale=0.5]{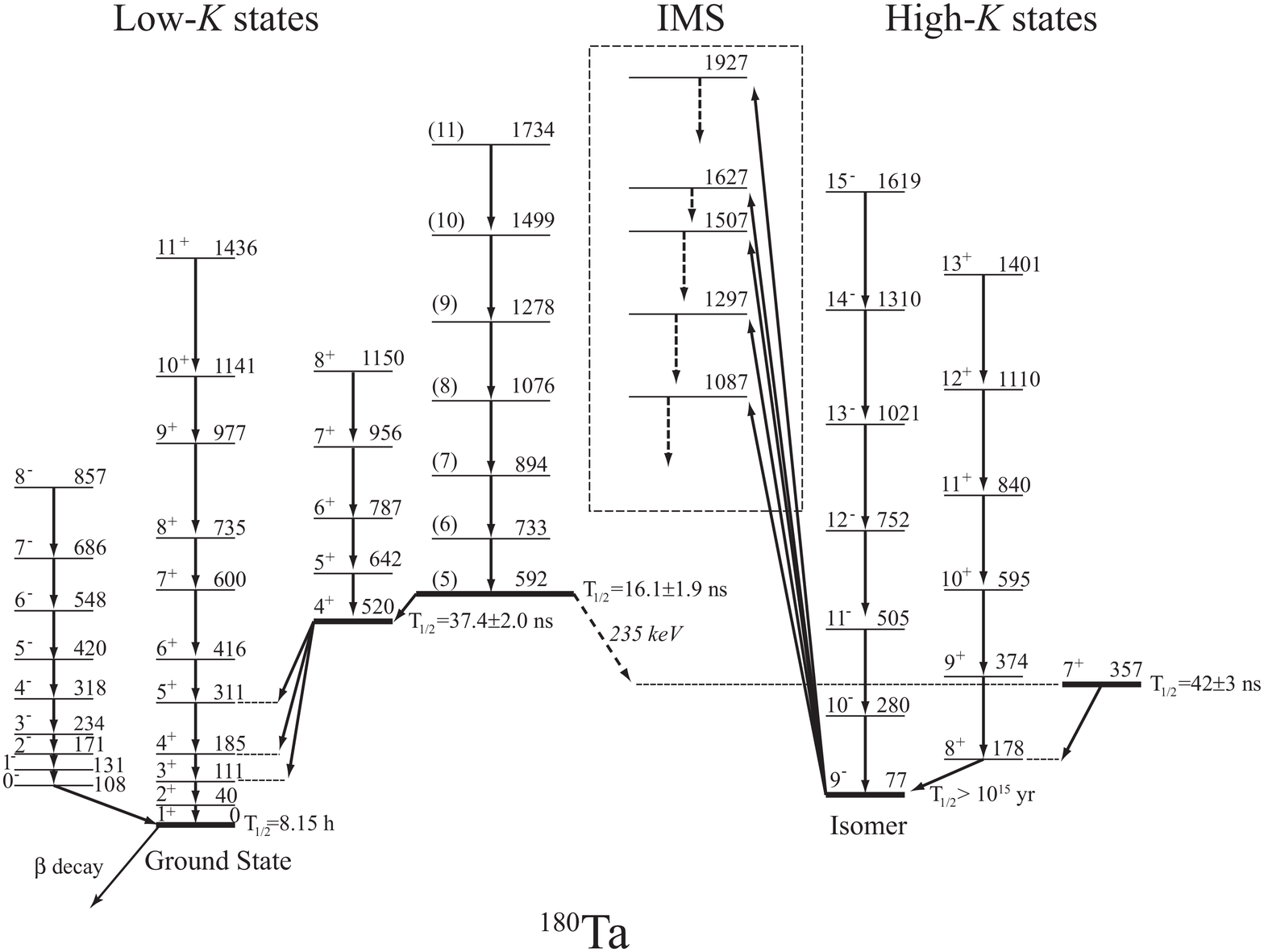}
  \caption{ 
    Partial level scheme of $^{180}$Ta. The excitation energies, spins, and
    parities of most levels 
are taken from the evaluated data \cite{Wu03}. The excitation energies of ($J$ = 5) band are taken from Refs.~\cite{Dra98,Sai99}. 
}
  \label{fig:level}
\end{figure*}

\begin{figure}[thbp] 
  \centering
  \includegraphics[scale=0.6]{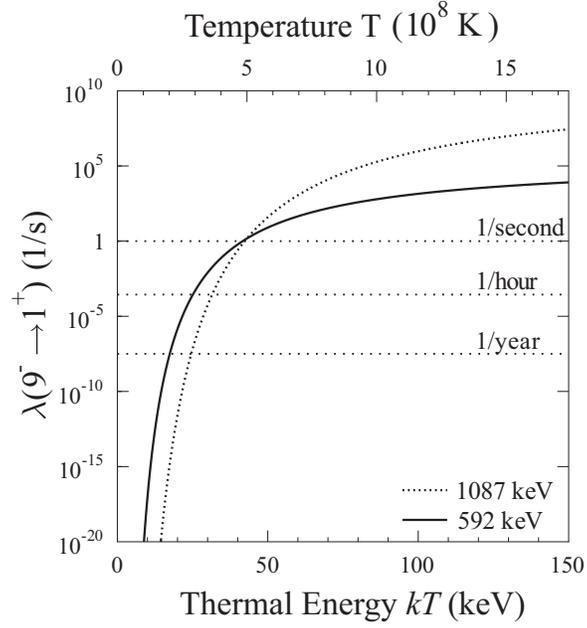}
  \caption{ 
    Reaction rates $\lambda^\ast$ under stellar
    conditions for the transition from the \hiK\ isomer to the
    \loK\ ground state in \taN .  
   Lines on this figure show contributions from the experimentally confirmed
    IMS at 1087\,keV (dotted line ) and the suggested IMS at 592\,keV
    (solid line). The relevant freeze-out region is
    between $4.4 \le T_8 \le 6.2$ where $T_8$ is the temperature in $10^8$\,K.  
}
  \label{fig:rates}
\end{figure}

\end{document}